# RELIABLE RESOURCE SELECTION IN GRID ENVIRONMENT

Rajesh Kumar Bawa[1] and Gaurav Sharma[2]

[1]Associate Professor, Deptt. of Computer Science, Punjabi University, Patiala, India
`rajesh_k_bawa@yahoo.com`
[2]Assistant Professor, Deptt. of Computer Engg., JMIT Radaur, Yamunanagar, India
`gauravsharma@jmit.ac.in`

*ABSTRACT*

*The primary concern in area of computational grid is security and resources. Most of the existing grids address this problem by authenticating the users, hosts and their interactions in an appropriate manner. A secured system is compulsory for the efficient utilization of grid services. The high degree of strangeness has been identified as the problem factors in the secured selection of grid. Without the assurance of a higher degree of trust relationship, competent resource selection and utilization cannot be achieved. In this paper we proposed an approach which is providing reliability and reputation aware security for resource selection in grid environment. In this approach, the self-protection capability and reputation weightage is utilized to obtain the Reliability Factor (RF) value. Therefore jobs are allocated to the resources that posses higher RF values. Extensive experimental evaluation shows that as higher trust and reliable nodes are selected the chances of failure decreased drastically.*

## 1. INTRODUCTION

Grid computing, most simply stated, is distributed computing taken to the next evolutionary level. The goal is to create the illusion of a simple yet large and powerful self managing virtual computer out of a large collection of connected heterogeneous systems sharing various combinations of resources. Grids enable the sharing, selection, and aggregation of a wide variety of resources including supercomputers, storage systems, data sources, and specialized devices that are geographically distributed and owned by different organizations for solving large-scale computational and data intensive problems in science, engineering, and commerce. Thus creating virtual organizations and enterprises as envisioned in as a temporary alliance of enterprises or organizations that come together to share resources and skills, core competencies, or resources in order to better respond to business opportunities or large-scale application processing requirements, and whose cooperation is supported by computer networks.

The processing speed of any computer is enormously increased. Even then it is not enough to satisfy the needs of researchers, scientist and for very high computation projects. Grid computing provides huge processing power in a distributed environment with variety of resources. It is a dynamic environment where the state and availability of the resource changes from time to time. Hence, the resources are to be allocated to clients based on the dynamic status information of the resources. The resources in Grid are shared in a flexible, coordinated and secured manner. Security is one of the prime issue in grid computing. In grid applications the security is in the terms of that the users should have reliable transactions. The reliability of any transaction is the probability of successful running or completion of a given task [11]. Grid computing and its related technologies will only be adopted by users, if they are confident that their data and privacy are secured and the system must be as scalable, robust and reliable as of their own in their places. Allowing reliable transactions plays a vital role in grid computing. Resources and security guarantee are the two fundamental requirements in Grid applications. Coordinated resource sharing and problem resolving in dynamic, multi institutional virtual organizations are the actual and specific problems which underlies the concept of grid. The sharing that we are concerned with is not primarily file exchange but rather direct access to computers, software, data, and other resources. To support scheduling and application execution in the context of the Grid, a Grid Resource Broker is desirable. Grid Resource Brokering is defined as the process of making scheduling decisions involving resources over multiple administrative domains. Grid Resource Discovery and Selection is an essential and crucial part of grid resource brokering, which provides adequate available resources for other grid resources management. Resource discovery is a previous process before resource selection. The goal of resource discovery is to identify a list of authenticated resource objects that are available for resource users. Once the list of possible target resource objects is known, the second phase is to select those resource objects that best suit the constraints and conditions imposed by the user, such as CPU usage, RAM available or disk storage. Resource selection involves a set of factors, such as application minimal requirements, application run time, and resource access policies [1, 2]. In addition, resource selection must consider uncertainties associated with each resource and answer questions related to resource reliability, prediction error probability, and cost error probability to access a resource. Grid technologies allows resource sharing among several entities, but selecting the most appropriate and secure resource to run a specific job remains one of its main problems. Most of the Grid applications involve very large data bases with highly secured data. Security requires the three fundamental services: authentication, authorization, and encryption. A grid resource must be authenticated before any checks can be done as to whether or not any requested access or operation is allowed within the grid. Once the grid resources have been authenticated within the grid, the grid user can be granted certain rights to access a grid resource. But within the grid application the one who uses the resource also needs reliable and secure services. So there is a need of reliable system which ensures a level of robustness against malicious nodes [12]. Users are able to submit jobs to remote resources and typically have no explicit control over the resources themselves. Therefore, mutually users and resources can be viewed as independent agents, having control of their own behavior. Since an individual cannot forecast the response of another to changing situations, this autonomy provides rise to inherent in security. So a better security mechanism is essential and crucial for secure and reliable communication in grid.

## 2. RELATED WORK

Trust has been addressed at different levels by many researchers. Several trust models that are related to our work has been proposed for integration into grid computing systems. Farag.A.Azzedin and Maheswaran.M describe the behavior trust model is where the trust is computed as a combination of direct experiences and reputation. Trust based on only subjective knowledge does not perform well in a dynamic grid environment. The execution environment parameters like communication speed, work load greatly influence the success of jobs in grids which are not addressed in this paper [3,4]. An adaptive trust model based on reputation is presented by Li Xiong, Ling Liu. The trust model quantifies and compares the trustworthiness of peers based on a transaction feedback system. The authors address about peer to peer trust and misbehaving of peers. Here, trust is evaluated based on community reputations. But the trust computed solely on recommendation mechanism is inaccurate and inefficient. In this paper the proposed method considers transaction context factor to select the resources but has not explored any mechanism to identify false feedbacks and honesty of the recommender [8]. A vector model for developing trust has been proposed by Indrajit Ray and Sudip Chakraborty. The trust rating between a trustor and a trustee is determined by the component values of experience, knowledge and recommendation. The normalized trust rating is given by the normalization of the trust policy vector and simple trust relationships. The model incorporates trust dynamics, change of trust and distrust with time [6]. Trust in virtual communities is discussed by Abdul-Rahman.A and Hailes.S. The reputation based trust defined is agent and context specific. The grade of outcome of an experience is given in terms of ordered set representing 'very good', 'good', 'bad', 'very bad'. The proposed model has uncertainty in the information if more than one value is returned as experience and therefore it is not robust to malicious encounters and risky environment s[5]. Xudong Ni , Junzhou Luo, a trust aware access control in service oriented grids is presented. Trust is applied to access control and the trust value is modified according to the increasing times of service. Trust is also used to determine the authorization of grid users and is a basic parameter of access control policy decision. Trust is computed with reference to the number of times a transaction is successful between a trustor and a trustee. This model has not considered time of past interactions. As trust changes over time, computing trust values based on time helps to find a more appropriate resource in the dynamic grid environment [7]. Trust as applied to scheduling in commercial grids is implemented by Thamaraiselvi.S, Balakrishnan.P, Kumar.R, Rajendar.K. The proposed trust model evaluates trust based on affordability, success rate and bandwidth. User jobs are submitted to resources with higher trust values. The feedback values specified doesn't consider the credibility of the recommender. Hence, there is a chance that a malicious resource can be chosen by the trust model which could harm the user's job running on the Grid platform [9]. Wenbo Mao,Fai Yan,Chunrun Chen suggests a Trust model which uses trusted computing model as the basic model. The behavior trust model is built upon Trusted platform module. This paper demands that the Grid resource selection can be improved by merging trusted computing in the behavior model [13]. According to Bendahmane et. al security becomes necessary to provide authentication, authorization, resource protection, secure communication, data confidentiality, data integrity, trust policies management, user key and credential management, service protection, and network security. They provides an insight classification of the different mechanisms and methods of security in grid computing environment, and also discusses the solutions for each category and its limitation in order to define the real problems in grid computing security. In their approach, the security solutions is categorize into Resources Level, Service Level, Authentication & Authorization Level, Information Level, and Management Level Solutions [14]. Vijayakumar et. al have proposed an approach, which intends to offer trust and reputation aware security for resource selection in grid computing. Based on the calculated trust factor (TF) value of each entity the incoming jobs have securely allocated to the entity. The effective and competent exploitation of grid computing services needs sophisticated and secured resource management systems. The wide range of selection and the high degree of strangeness leads to the problem in secured selection of grid. Without the assurance of a higher degree of confidence relationship, efficient resource allocation and utilization cannot be attained [15]. Anthony R. Metke and Randy L. Ekl proposed a paper which aims at to increase overall system efficiency and reliability. According to them, much of the technology currently in use by the grid is outdated and in many cases unreliable. There have been three major blackouts in the past ten years. The reliance on old technology leads to inefficient systems, costing unnecessary money to the utilities, consumers, and taxpayers. To upgrade the grid, and to operate an improved grid, will require significant dependence on distributed intelligence and broadband communication capabilities. The access and communications capabilities require the latest in proven security technology for extremely large, wide-area communications networks. Their paper discusses key security technologies for a smart grid system, including public key infrastructures and trusted computing [16].

### 3. PROPOSED APPROACH

This section explains our proposed approach of resource selection designed for safe scheduling of independent and individual jobs to grid sites. The scale of resources and the strangeness of entities cause difficulties in the process of resource selection. So a proper safety mechanism is required for grid resources, because if resource is affected by virus or some malicious code then it will not execute the application safely and degrades the performance of the grid .In this paper resource reputation is calculated which is based upon the user feedback. The approach aims for secure scheduling of incoming jobs based on the Reliability Factor to available resource sites. The Reliability Factor (RF) value of each resource site is calculated through its self-protection capability and reputation weightage obtained from user community on its past behavior. Two necessary assumptions are made below: (a) all resource sites have prior agreements to participate in the Grid operations; and (b) the Grid sites truthfully report their self-protection capability to Grid organization manager (GOM).

### 3.1 Self-Protection Capability

The grid organization manager maintains the self protection capability of all entities in a grid organization. Every so often each entity reports its self-protection capability trustfully and honestly to the GOM. The self protection capability of an entity is calculated by aggregating the values of the below mentioned security factors. The value of these factors differs in the range between 0 and 1.
• Anti-spyware (AS): - the ability of the resource to monitors cookies, adware, spyware and other types of malicious code invading the computer.

• Anti-virus Capabilities(AVC): - The ability of the resource to defend against viruses and malicious codes.

• Firewall Capabilities (FC): - The ability to protect the node from other network accesses.

• Authentication Mechanism (AM): - The ability of the mechanism to verify an identity claimed by or for a system security.

• Backup Facility (BF): - The ability of the resource to provide backup facility of securely storing the data needed for the execution of job.

• Network analyzer (NA): - The ability of the resource to Provide detailed statistics for current and recent activity on the network to enhance protection against malicious activity.

• Internet Protocol Security (IPS): -The ability of the resource to provide integrity and confidentiality by using encryption.

Based on their contribution to security, a weightage is given to all the security factors and as a final point aggregated to compute the self-protection capability. The weightage assigned to the security factors are listed in Table1.

Table:1. Security Factors

| Security type | Value Associated |
| --- | --- |
| Anti-spyware | 0.82 |
| Antivirus capabilities | 0.85 |
| Firewall capabilities | 0.9 |
| Authentication mechanism | 0.8 |
| Backup Facility | 0.7 |
| Network analyzer | 0.6 |
| Internet Protocol Security (IPSec) | 0.75 |

The self-protection capability is calculated using the following formula

$$SPC = \sum_{i=1}^{n} \frac{A_i}{n}$$

Where *n* is the total number of factors, *A(i)* value of the factor.

## 3.2. Resource Reputation Computation

Since reputation is a multi-faceted concept [10], it has many aspects for instance truthfulness, honesty and so on. Reputation weightage is calculated via the feedback provided by the user community about their previous experiences. After the usage, users will provide feedback on the attributes to the GOM based on their experience. The feedback is a value in the range between 0 and 1. A nodes's feedback from all the users has aggregated. The GOM in grid organization maintains the reputation weightage of all nodes. The security attributes considered for the reputation are as follows.

- Node Consistency (NC): - The ability of the node to perform its assigned task with same efficiency for a specified period of time.

- Node independence (NI):- The node executes the job independently without interfering of local jobs running on the system. It means node does not share its resources like CPU, RAM with local jobs running on the system.

- Node Truthfulness (NT): - The ability of the Node to ensure that the data is protected from unauthorized modifications.

- Node Privacy (NP): The application which is submitted to node is kept confidential and does not breach privacy of the provider.

- Node Utilization (NU): It is defined as the percentage of processing power allotted to user's task out of the total processing power available at a selected node.

$$\text{Node Utilization} = \frac{\sum_{i=1}^{n} UPC_i}{TPC}$$

Where,
UPCi is the Utilized Power Compute for the ith task
TPC is the Total Power Compute of the selected resource Ri.

- Node Reliability (NR): Reliability of a node is defined as the number of jobs executed successfully to the total number of jobs submitted to the node. The reliability of a node R for n jobs is given as,

$$NR = \frac{\text{No. of job successful excuted}}{\text{Total No. of jobs submitted}}$$

- Node Authorization (NA): - Refers to the process of granting privileges to processes and, ultimately, users.

We estimate the feedback value of the node based upon the above factors like node consistency, Node independence, Node Truthfulness, Node Privacy and all. The range of these factors lies between 0 to1.The aggregated feedback of all the attributes of an node is represented as a Reputation Weigthage ( RW) and it is calculated using the following formula:

$$RW(Ea) = \sum_{i=1}^{n} \frac{Ei}{n}$$

$E_i$ = The value of all the feedback factors (between 0 to1)
$N$ = The total number of factors

The reliability factor ($RF$) of each node is calculated by utilizing the self-protection capability ($SPC$) and Reputation Weigthage ( RW).

$$RF(E_a) = \frac{SPC(E_a) + RW(E_a)}{2}$$

Higher RF value mean the resource having better security mechanism and positive feedback as compared to the other resources. Thus apply ranking mechanism and rank the resources according to their RF value, higher the RF value of the resource higher the resource rank. The highest rank value node is selected for job submission.

**The algorithm for Secure Resource Selection:-**

For

    Each node in Grid domain

1. *findout all the security factors available in the resources*
2. *Calculate the SPC value of the resource by computing factors associated value*
3. *Obtain RW by considering Feedback factor values from the users.*
4. *Calculate RF value of the resource.*

Repeat step 1 to 4 for all resource in the grid

5. *Rank the resources according to RF value.*
6. *Submit the job to the node having highest RF value.*

    End

## 4. EXPERIMENTAL RESULT

In this section, we first describe the experimental setup and present the analysis of our experimental results. The proposed algorithm is implemented in Java. The experimental setup consists of seven grid nodes and a Grid Organization Manager (GOM)

Table2- SPC factor of various nodes

| Nodes | AS | AVC | FC | AM | BF | NA | IPS |
|---|---|---|---|---|---|---|---|
| N1 | 0.25 | 0.54 | 0.66 | 0.6 | 0.6 | 0.7 | 0.6 |
| N2 | 0.2 | 0.5 | 0.7 | 0.59 | 0.59 | 0.8 | 0.87 |
| N3 | 0.6 | 0.37 | 0.89 | 0.51 | 0.67 | 0.73 | 0.79 |
| N4 | 0.15 | 0.21 | 0.45 | 0.57 | 0.39 | 0.23 | 0.38 |
| N5 | 0.145 | 0.7725 | 0.7775 | 0.675 | 0.7075 | 0.675 | 0.7 |
| N6 | 0.6 | 0.37 | 0.89 | 0.51 | 0.67 | 0.73 | 0.79 |
| N7 | 0.51 | 0.42 | 0.5 | 0.56 | 0.7 | 0.4 | 0.61 |

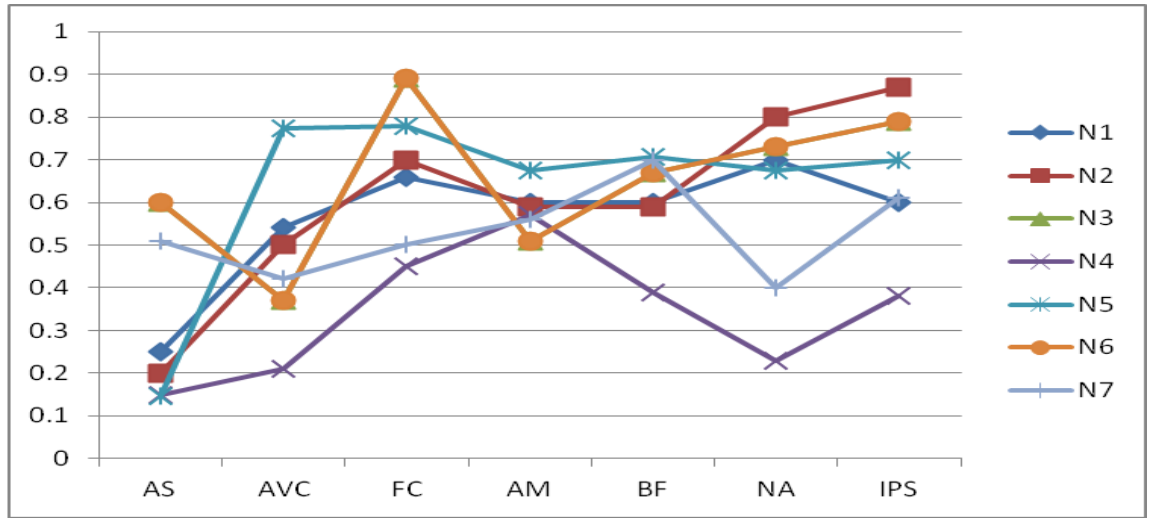

Fig. 1- SPC capabilities of each node

At first, the users submit their jobs to GOM. The GOM will calculate the trust factor value of all the nodes based on their Self protection capability and Reputation weightage. A node with high trust factor value is selected for the execution of current job. The GOM will inform the user with the selected node for their job execution. After the completion of job, the user is asked to provide feedback about the node on some security attributes. The selected node has provided high security for the job execution. The self protection capability of all the nodes is updated by the GOM in a periodical manner. The reputation weightage is frequently updated for all the nodes based on the feedback value from user communities. The security factors utilized for determining the self-protection capability of the seven grid nodes are enlisted along with their respective values in Table2. The security attributes that eventually aid in the estimation of reputation weightage with their respective values are as well listed subsequently in Table 3. The RF value of the nodes is calculated by considering their corresponding SPC and RW value which is also listed in table 4. Based upon RF value we provide the ranking mechanism.

Table 3- Reputation Weightage

| Node | NC | NI | NT | NP | NP | NU | NR | NA |
|------|------|-------|------|-------|------|------|------|------|
| N1 | 0.25 | 0.29 | 0.3 | 0.35 | 0.4 | 0.31 | 0.12 | 0.35 |
| N2 | 0.68 | 0.69 | 1 | 0.4 | 0.1 | 0.35 | 0.21 | 0.7 |
| N3 | 0.6 | 0.7 | 0.8 | 0.25 | 0 | 0.21 | 0.6 | 0.58 |
| N4 | 0.71 | 0.77 | 0.85 | 0.52 | 0.23 | 0.58 | 0.15 | 0.67 |
| N5 | 0.46 | 0.463 | 0.47 | 0.43 | 0.3 | 0.44 | 0.19 | 0.44 |
| N6 | 0.54 | 0.725 | 0.75 | 0.465 | 0.4 | 0.5 | 0.5 | 0.6 |
| N7 | 0.75 | 0.8 | 0.9 | 0.28 | 0.15 | 0.32 | 0.51 | 0.55 |

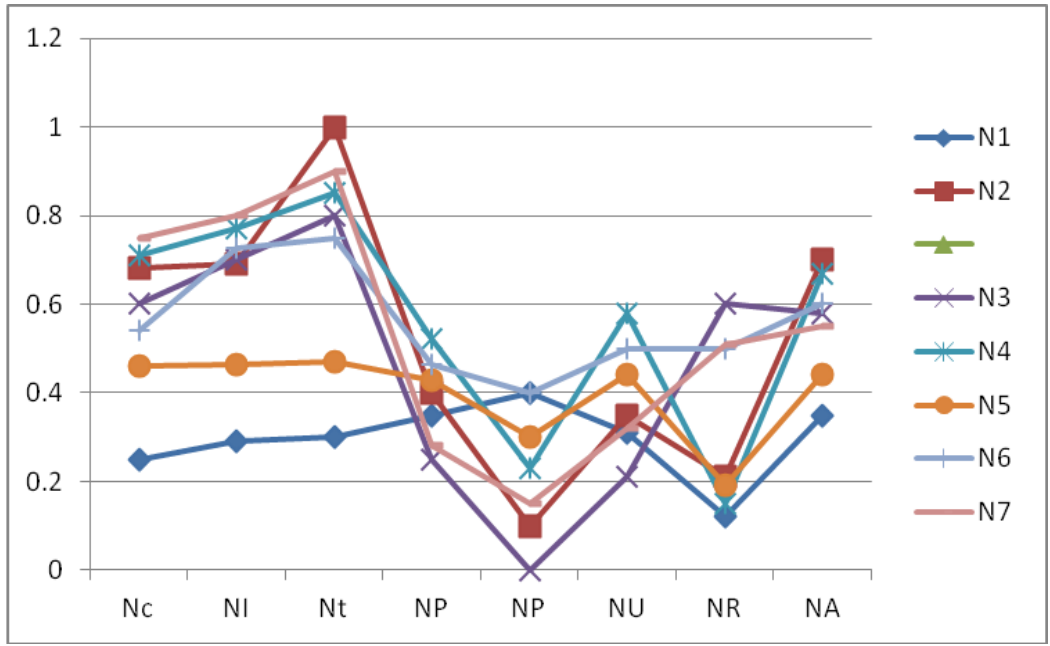

Fig. 2: Security Attributes of each node

Table 4-Calculated SPC, RW &RF value

| Nodes | SPC Value | RW Value | RF value |
|---|---|---|---|
| N1 | .564 | .281 | .422 |
| N2 | .607 | .516 | .561 |
| N3 | .651 | .467 | .599 |
| N4 | .34 | .565 | .452 |
| N5 | .636 | .399 | .517 |
| N6 | .654 | .56 | .607 |
| N7 | .528 | .532 | .53 |

Higher RF value means the resource having better security mechanism and positive feedback as compared to the other resources. In our experiment we submitted 1000 jobs to above 7 nodes and calculate the failure rates.

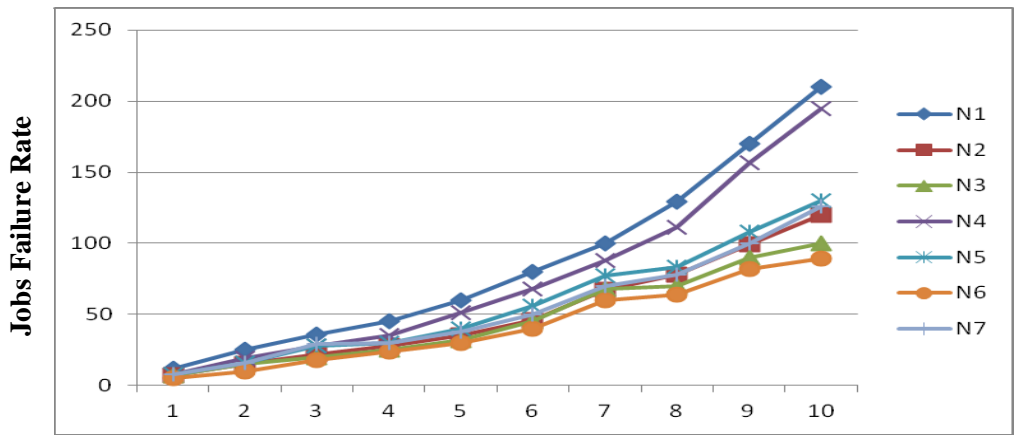

Figure 3: Failure Rate Comparison

It was observed from Fig3 that in the beginning of jobs submission the failure rate is almost same but as the number of jobs increased the failure rate is increased drastically. It was also observed that the Nodes having higher RF value encounters lesser failures as shown in fig3.The nodeN6 has highest RF value and lowest failure rate**.** So the Node N6 is selected for job submission which is more reputed and more secure.

## 5. CONCLUSION

Computational Grids are quickly rising as a practical means by which to execute new science and develop new applications. The effective and efficient exploitation of Grid computing facilities needs highly advanced and protected resource management systems. Efficient resource sharing and accessing cannot go without the assurance of high trustworthiness. Reputation mechanisms provide a way for building trust through social control using community based feedback about previous experiences. In this paper, we have proposed a secured approach for the users in selecting the correct resource meant for their job execution. The proposed approach joint both trust and reputation to provide security for resource selection mechanism. Our approach aggregates several security related attributes for both self protection capability and reputation into numerical values, which can be easily applied to calculate the Trust factor of grid entity.

**Authors**

**Dr. Rajesh K. Bawa** holds Master's and Ph.D degree in the area of Numerical Computing From IIT Kanpur INDIA. Presently he is Associate Professor at Department of Computer science Punjabi University, Patiala,INDIA.His present area of interest is Parallel and Scientific Computation and has published many research papers in journals of reputed publishers and Also associated with review and editing work with many journals.
Email: rajesh_k_bawa@yahoo.com

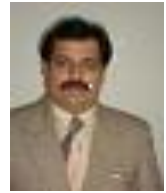

**Gaurav Sharma** received his B. Sc.and M. Sc.degrees in computer science from the Kurukshetra University, Kurukshetra in 1999 and 2001, respectively. Also, he holds the degree of M.Tech in Computer Science from Guru Jambeshwar University, Hisar.He is currently a Ph. D.candidate in Department of Computer Science, Punjabi University, Patiala. Also, he is working as Assistant Professor in Engineering college, JMIT Radaur. His research interests include grid computing, Software engineering, and artificial intelligence.
E-mail: gauravsharma@jmit.ac.in (Corresponding author)

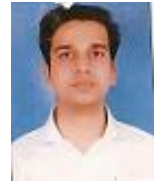